 \documentclass[final,5p,times,twocolumn,square,comma,compress]{elsarticle}

\usepackage{amssymb}
\usepackage{lipsum}

\usepackage[colorlinks,
bookmarks=true,
linkcolor=blue,
urlcolor=blue,
anchorcolor=black,
citecolor=blue
]{hyperref}
\usepackage{graphicx}
\usepackage{dcolumn}
\usepackage{bm}
\usepackage{subcaption}
\usepackage{amsmath}

\journal{Physics Letters B}

\begin{document}

\begin{frontmatter}



\title{Investigation of $T_{cs0}^{*}(2870)^{0}$ in $pp$ collisions at $\sqrt{s}$ = 7 TeV with the PACIAE model}


\author[first]{Qiang Wang}
\affiliation[first]{organization={Key Laboratory of Quark and Lepton Physics (MOE) and Institute of
Particle Physics, Central China Normal University},
            city={Wuhan},
            postcode={430079},
            country={China}}

\author[second]{Zhi-Lei She}
\ead{shezhilei@cug.edu.cn}
\affiliation[second]{organization={School of Mathematics and Statistics, Wuhan Textile
University},
            city={Wuhan},
            postcode={430200},
            country={China}}

\author[first,third]{An-Ke Lei}
\affiliation[third]{organization={School of Physics and Electronic Science, Guizhou Normal University},
            city={Guiyang},
            postcode={550025},
            country={China}}

\author[first]{Dai-Mei Zhou}
\ead{zhoudm@mail.ccnu.edu.cn}

\author[fourth]{Wen-Chao Zhang}
\affiliation[fourth]{organization={School of Physics and Information Technology, Shaanxi Normal
University},
            city={Xi'an},
            postcode={710119},
            country={China}}

\author[fourth]{Hua Zheng}

\author[fifth]{Yu-Liang Yan}
\ead{yuliang86@yeah.net}
\affiliation[fifth]{organization={China Institute of Atomic Energy, P. O. Box 275 (10)},
            city={Beijing},
            postcode={102413},
            country={China}}

\author[first,fifth]{Ben-Hao Sa}
\ead{sabhliuym35@qq.con}

\begin{abstract}
We have used the parton and hadron cascade model PACIAE together with the
Dynamically Constrained Phase-space Coalescence model (DCPC) to study the
$T_{cs0}^{*}(2870)^{0}$ production in $pp$ collision at $\sqrt{s}$ = 7 TeV, following the LHCb observation of $T_{cs0}^{*}(2870)^{0}$ in the
$B^{-}\to D^{-}D^{0}K^{0}_{S}$ decays in $pp$ collisions at
$\sqrt{s}$ = 7, 8, and 13 TeV [PRL 134(2025)101901].
The final hadronic states of the $pp$ collisions at $\sqrt{s}$ = 7 TeV are
first simulated by the PACIAE model. Four
sets of $T_{cs0}^{*}(2870)^{0}$ candidates are then recombined by the DCPC
model using the constituent meson pair of $D^{0}K^{0}_{S}$, $D^{+}K^{-}$, $D^{*+}K^{*-}$, and $D^{*0} \bar{K}^{*0}$ based on the above
simulated final hadronic states, respectively. We calculate their rapidity
distributions, transverse momentum spectra, and angular distribution between
the two component mesons, as well as angular distribution between $D$
component meson and $T_{cs0}^{*}(2870)^{0}$. Our results show that the yields
of four $T_{cs0}^{*}(2870)^{0}$ candidates follow the magnitude order of $D^{0}K^{0}_{S}$ $>$ $D^{+}K^{-}$ $>$ $D^{*+}K^{*-}$ $\sim$ $D^{*0} \bar{K}^{*0}$. Similar ordering behavior is also observed in the
aforementioned distributions.
\end{abstract}







\end{frontmatter}




\section{Introduction}
\label{introduction}
The existence of exotic states containing more than three quarks is
allowed by the quark model and quantum chromodynamics (QCD) theory~\cite{Gell-Mann:1964ewy,Zweig:1964ruk,Jaffe:1976ig}. The first exotic hadron
candidate X(3872) (or $\chi_{c 1}(3872)$) was experimentally observed by the
Belle Collaboration in 2003~\cite{Belle:2003nnu}. Since then more and more
exotic states have been confirmed experimentally \cite{Chen:2022asf,Johnson:2024omq},
including $T_{cc}(3875)^{+}$ \cite{LHCb:2021vvq,LHCb:2021auc}, $T_{cc\bar{c}\bar{c}}(6900)^{0}$ \cite{LHCb:2020bwg,ATLAS:2023bft,CMS:2023owd}, etc. In 2020, the LHCb
Collaboration reported two new exotic states in the
$D^{+}K^{-}$ final state of the $B^{-}\to D^{-}D^{+}K^{-}$ decays: $T_{cs0}^{*}(2870)^{0}$
and $T_{cs1}^{*}(2900)^{0}$, with quark content of $cs\bar{u}\bar{d}$ and
with quantum number of $J^{P}=0^{+}$ and $1^{-}$,
respectively~\cite{LHCb:2020pxc,LHCb:2020bls}.
They are the first two manifestly observed open-charm tetraquark exotic states
containing one single charm quark, revealing a new horizon for understanding
the mechanism of quark confinement. In 2025, the LHCb collaboration confirmed
the existence of $T_{cs0}^{*}(2870)^{0}$ in the $D^{0}K^{0}_{S}$ invariant
mass spectrum from the new decay mode of $B^{-}\to D^{-}D^{0}K^{0}_{S}$ in $pp$
collisions at $\sqrt{s}$ = 7, 8, and 13 TeV. Unfortunately, there is no evidence for
$T_{cs1}^{*}(2900)^{0}$ in the mass
spectrum of $D^{0}K^{0}_{S}$ in this channel~\cite{LHCb:2024xyx}. These
experimental observations will deepen our understanding of the nature of
$T_{cs0}^{*}(2870)^{0}$.
	
In analogous to other exotic states, the physical structures of
$T_{cs0}^{*}(2870)^{0}$ and $T_{cs1}^{*}(2900)^{0}$ are controversial, and
their nature is still unclear. In Refs.~\cite{Karliner:2020vsi,
He:2020jna,Guo:2021mja,Ozdem:2022ydv,Wang:2020xyc,Agaev:2021knl,Yang:2021izl,
Wang:2020prk,Jiang:2023rcn} they have been suggested as compact tetraquark
states. But
they were interpreted as hadronic molecular states in Refs.~\cite{Hu:2020mxp,Liu:2020nil,Kong:2021ohg,Wang:2021lwy,Xiao:2020ltm,Chen:2020aos,Chen:2021xlu,
Huang:2020ptc,Molina:2010tx,Molina:2020hde,Jin:2016cpv}.
The relative momentum between the hadrons in a hadronic molecule is much smaller than the hadron mass, making it suitable for analysis with non-relativistic methods ~\cite{Jin:2016vjn}.   
The nature of exotic states as hadronic molecules can be well accounted for by the coalescence mechanism~\cite{Wu:2022wgn}. Therefore,
in this work, we assume $T_{cs0}^{*}(2870)^{0}$ produced in the final
hadronic states (FHS) is a
hadronic molecular state and employ the parton and hadron cascade model
PACIAE~\cite{Lei:2024kam} together with the Dynamically Constrained
Phase-space Coalescence model (DCPC~\cite{Yan:2011fq}; PACIAE + DCPC) to study
its nature and properties.

The PACIAE + DCPC model has been successfully used to describe the production of light nuclei (such
as deuterons and hypertritons)  in heavy-ion collisions at
various center-of-mass energies \cite{Yan:2011fq,She:2020qyp}. It has also
been applied to describe the production of exotic states, e.g.,
X(3872) \cite{Xu:2021drf,She:2024cit,Wu:2023aim}, $P_c$ states \cite{Chen:2021ifb},
$Z_c^{\pm}$(3900) \cite{Zhang:2020vfv}, X(2370) \cite{Cao:2024mfn}, and
G(3900) \cite{Cao:2025uvw} in nuclear collisions.

In this letter, the PACIAE model is first employed to simulate the final
hadronic states (FHS) in $pp$ collisions at $\sqrt{s}$ = 7 TeV.
The $T_{cs0}^{*}(2870)^{0}$ molecular state is then coalesced in this FHS
with the component meson pair of $D^0K^0_S$ and $D^+K^-$, by the DCPC model. These two component meson pairs correspond
to the $T_{cs0}^{*}(2870)^{0}$ candidates observed by LHCb in decay channels
of $B^-\to D^-D^0K^0_S$ and $B^-\to D^-D^+K^-$, respectively. The four sets
of $T_{cs0}^{*}(2870)^{0}$ candidates composed of $D^0K^0_S$, $D^+K^-$, $D^{*+}K^{*-}$, and $D^{*0} \bar{K}^{*0}$ have the same quark content of $cs\bar{u}\bar{d}$. We
calculate their rapidity ($y$) distributions, transverse momentum ($p_T$)
spectra. Then we compute the angular distribution between two component mesons of $D$ and $K$
($D^{0}$ and $K^{0}_{S}$, $D^{+}$ and $K^{-}$), as well as the angular distribution
between $D$ ($D^{0}$ or $D^{+}$) component meson and
$T_{cs0}^{*}(2870)^{0}$. Our results show that the yield of
$T_{cs0}^{*}(2870)^{0}$ composed of $D^{0}K^{0}_{S}$ is larger than the one composed
of $D^{+}K^{-}$,
while the yields of $T_{cs0}^{*}(2870)^{0}$
composed of $D^{*+}K^{*-}$ and  $D^{*0} \bar{K}^{*0}$ are lower than the former two and almost equal.
Similar ordering behavior is also found in all the aforementioned distributions.

\section{Methodology}
\label{Methodology}
\subsection{PACIAE model generating final hadronic state}
The parton and hadron cascade model PACIAE~\cite{Sa:2011ye,Lei:2023srp} based
on PYTHIA~\cite{Sjostrand:2006za}, can be used to describe high-energy
elementary particle collisions as well as nucleus-nucleus collisions. It
contains four stages of the parton initialization, the partonic
rescattering, the hadronization, and the hadronic rescattering. The latest version
of PACIAE 4.0 \cite{Lei:2024kam} is based on
PYTHIA 8.3~\cite{10.21468/SciPostPhysCodeb.8}. In this letter, we utilize it
to generate the final hadronic state in $pp$ collisions at $\sqrt{s}$=7 TeV.
	
In the parton initialization stage, each nucleon-nucleon (hadron-hadron)
collision is executed by PYTHIA model \cite{10.21468/SciPostPhysCodeb.8} with
presetting the string fragmentation turn-off temporarily. The obtained
parton initial state is produced by QCD hard scattering and associated
initial and final QCD radiation. Then the gluons split and the
energetic quarks (antiquarks) deexcitation is implemented. In the next stage
of partonic rescattering, the leading-order perturbative quantum chromodynamics
(LO-pQCD) parton-parton scattering cross section \cite{Combridge:1977dm} is
employed to supply the final partonic state (FPS) composed of abundant quarks
and antiquarks with their four-coordinate and four-momentum. The
hadronization process is then achieved via the Lund string fragmentation
mechanism and/or the phenomenological coalescence model \cite{Lei:2024kam},
and the former is chosen in this letter. Eventually, the final
hadronic state (FHS) is produced after the hadronic rescattering.
This FHS is composed of numerous hadrons with their four-coordinate and four-momentum.
In the process of hadronic rescattering, there are hundreds of
inelastic reaction channels of light hadrons, and more inelastic reactions involving the heavy hadrons are included \cite{Lei:2024kam}.
Here the reaction channels of $D$ and $K$ mesons are listed as an example:

\begin{equation*}
    \begin{aligned}
        D + \pi        &\rightleftharpoons  D^* + \rho,            &\quad
        D^* + \pi      &\rightleftharpoons  D + \rho,              \\
        D + N          &\rightleftharpoons  D^* + N,               &\quad
        D^* + \bar{N}  &\rightleftharpoons  D + \bar{N},           \\
        \bar K N       &\rightleftharpoons  \pi Y,                 &\quad
        K \bar{N}      &\rightleftharpoons  \pi \bar{Y},           \\
        \bar{K} Y      &\rightleftharpoons  \pi \Xi,               &\quad
        K \bar{Y}      &\rightleftharpoons  \pi \bar{\Xi},         \\
        \bar{K} N      &\rightleftharpoons  K \Xi,                 &\quad
        K \bar{N}      &\rightleftharpoons  \bar{K} \bar{\Xi},     \\
        K \bar{\Xi}    &\rightleftharpoons  \pi \bar{\Omega},      &\quad
        \bar{K} \Xi    &\rightleftharpoons  \pi \Omega,            
    \end{aligned}
\end{equation*}
where $ Y $ stands for $ \Lambda $ or $ \Sigma $.

\subsection{DCPC model recombining $T_{cs0}^{*}(2870)^{0}$ candidates}
The DCPC model is based on the quantum statistical mechanics~\cite{Thermo_Cambri,Thermo_North}, where the yield of $N$-particle cluster is defined as
\begin{equation}
	Y_1 = \int ... \int_{E_a \le H \le E_b } \frac{ \mathrm{d}\vec{r}_1 \mathrm{d}\vec{p}_1 ...
       \mathrm{d}\vec{r}_N \mathrm{d}\vec{p}_N }{h^{3N} } ,
\end{equation}
where $E_a$ and $E_b$ denote the lower and upper energy thresholds of the
cluster. The $\vec{r}_i$ and $\vec{p}_i$ are, respectively, the
coordinate and momentum of the $i$-th particle. In order for the cluster to 
exist naturally, some dynamical constraints are introduced in the DCPC
model. Thus the yield of $T_{cs0}^{*}(2870)^{0}$ composed of $D$ and $K$
component meson pair, for instance, is calculated by
\begin{equation}
	Y_{T_{cs0}^{*}(2870)^{0}} = \int ... \int \delta_{12} \frac{ \mathrm{d}\vec{r}_1 \mathrm{d}\vec{p}_1  \mathrm{d}\vec{r}_2 \mathrm{d}\vec{p}_2 }{h^{6} },
\end{equation}
where
\begin{equation}
	\delta_{12} =
	\begin{cases}
	      1  \quad \rm{if} \quad 1 \equiv D, 2 \equiv K,  \\
		\qquad m_0-\Delta m \le m_{\rm{inv} } \le m_0+\Delta m,  \\
		\qquad r_{i0} \le R_0, (i=1,2)  \\
		0  \quad \rm{otherwise}.
	\end{cases}\label{eq:DCPC_constr_1_ENG}
\end{equation}
In Eq.(\ref{eq:DCPC_constr_1_ENG}), $m_0$ ($\Delta m$) denotes the
mass (mass uncertainty parameter) of the cluster ($T_{cs0}^{*}(2870)^{0}$
candidate). $R_0$ and $r_{i0}$ refer to the radius of the cluster (a free
parameter) and the relative distance between the $i$-th component meson and
the center of mass of the cluster, respectively. The invariant mass $m_{\rm{inv}}$
is defined as
\begin{equation}
	m_{\rm{inv} } = \sqrt{(E_1+E_2)^2-( \vec{p}_1+\vec{p}_2 )^2},
\end{equation}
where $E_1$ ($E_2$) and $\vec{p}_1$ ($\vec{p}_2$) denote, respectively, the
energy and momentum of $D$ ($K$) component meson.		

\begin{figure}[!htbp]
    \centering
    \includegraphics[width=0.50\textwidth]{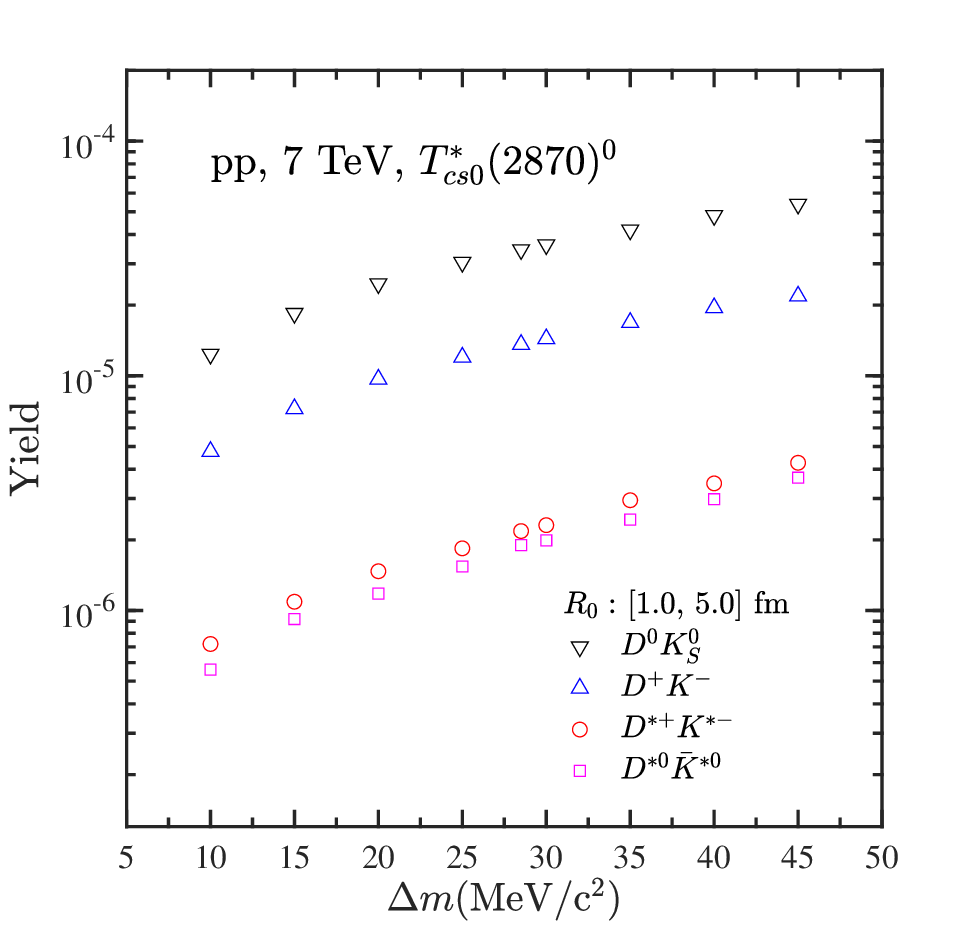}
    \caption{ \centering{The yield of $T_{cs0}^{*}(2870)^{0}$ recombined from $D^0 K^0_S$ (triangles-down), $D^+ K^-$ (triangles-up), $D^{*+} K^{*-}$ (circles), and $D^{*0} \bar{K}^{*0}$ (squares) varying with the parameter $\Delta m$ under the $R_0$ range of 1.0 fm $<R_0<$ 5.0 fm in $pp$ collisions at $\sqrt{s}$ = 7 TeV. }
    }\label{Tcs2870_display_yield_deltaM}
\end{figure}

\subsection{Calculation routine in DCPC}
We first construct a $T_{cs0}^{*}(2870)^{0}$ component meson list (consisted
of $K^-$, $K^0_S$, $K^{*-}$, $\bar{K}^{*0}$, $D^+$, $D^0$, $D^{*+}$, and $D^{*0}$) based on the hadron list in
the FHS generated by the PACIAE model. According to Eq.~(\ref{eq:DCPC_constr_1_ENG}) we then construct a two-layer circulations
(loops) over the component meson list to filter out any one of possible
candidate among the studied four sets of $T_{cs0}^{*}(2870)^{0}$ candidates
(composed of $D^{0}K^{0}_{S}$, $D^{+}K^{-}$, $D^{*+}K^{*-}$, and $D^{*0} \bar{K}^{*0}$). Here the
parameters in Eq.~(\ref{eq:DCPC_constr_1_ENG}) are set as follows: $m_0=2866$
MeV/$c^2$ (the experimentally observed mass peak of $T_{cs0}^{*}(2870)^{0}$
molecular state); $\Delta m$ equals to 28.5 MeV/$c^2$ (the value of the half decay width
for the $T_{cs0}^{*}(2870)^{0}$~\cite{ParticleDataGroup:2024cfk}). 
The dependence of $T_{cs0}^{*}(2870)^{0}$ yield on $\Delta m$ under the $R_0$ range of 1.0 fm $<R_0<$ 5.0 fm is shown in Fig \ref{Tcs2870_display_yield_deltaM}.
The yield of $T_{cs0}^{*}(2870)^{0}$ increases with $\Delta m$. 
The larger $\Delta m$ makes it easier for the cluster of $D$ and $K$ meson pair to satisfy the mass constraint condition in DCPC model and recombine the $T_{cs0}^{*}(2870)^{0}$.
The $R_0$ is assumed to be in the range of 1.0 fm $<R_0<$ 5.0 fm, or 1.0 fm  $<R_0<$ 3.0 fm, or
1.0 fm  $<R_0<$ 2.0 fm. The upper bound 2 fm is chosen because the spatial coalescence radius should be larger than the sum of radii for $D$ and $K$ meson pair~\cite{Cao:2024mfn}. 
The upper bound 5 fm is set due to the  assumed size of the particle-emitting source~\cite{Liu:2023uly}. 
To discuss the dependence of $T_{cs0}^{*}(2870)^{0}$ production on $R_0$, the upper bound 3 fm is also chosen.

\section{Results and discussion}
\label{RandD}
\begin{figure}[!htbp]
\centering
\includegraphics[width=0.50\textwidth]{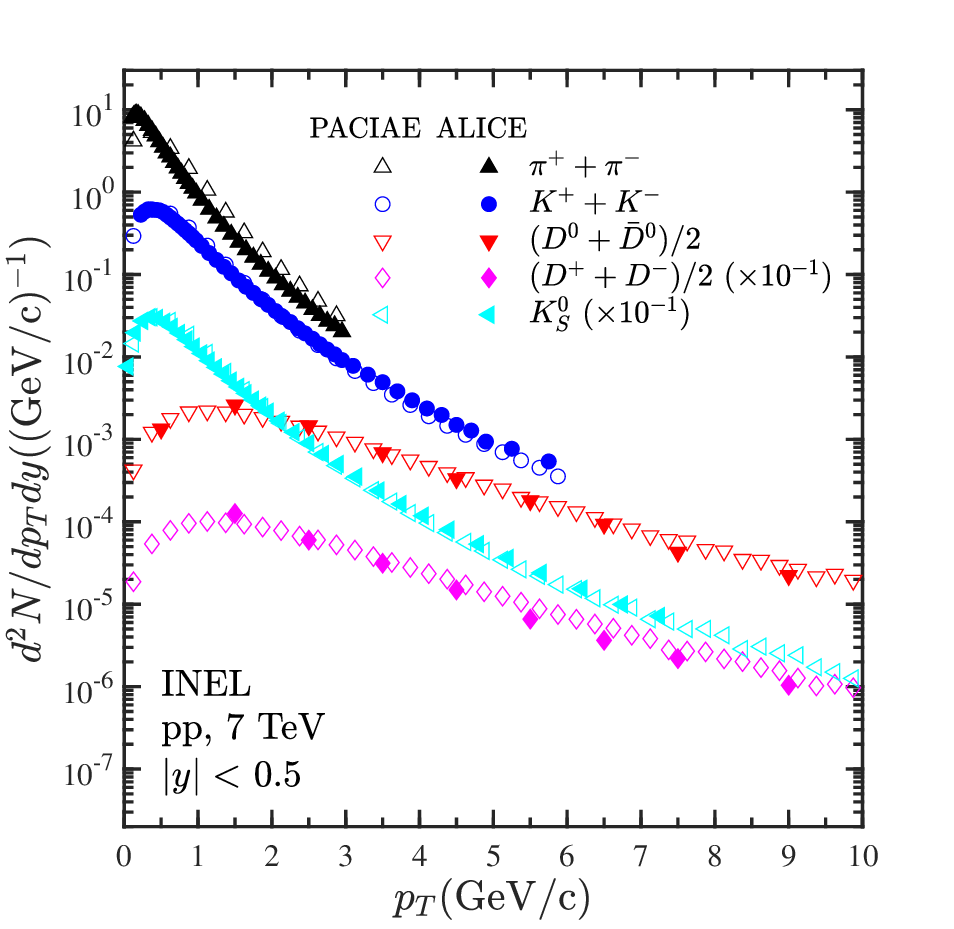}
\caption{ \centering{The simulated $p_T$ spectra and the
corresponding experimental data of $\pi^{+}+\pi^{-}$, $K^{+}+K^{-}$, $K^{0}_{S}$, $(D^{0}+\bar{D}^0)/2$ and $(D^{+}+D^{-})/2$. The empty markers refer to the results simulated by the PACIAE model, and the solid markers refer to the ALICE data~\cite{ALICE:2015ial,ALICE:2017olh,ALICE:2020jsh}. }
}\label{fig_Tcs2870_pi_K_D_dN_dpT}
\end{figure}

The PACIAE 4.0 model (with string fragmentation) is used to generate 800
million final hadronic states for the $pp$ collisions at $\sqrt{s}$ = 7 TeV.	
Here we adjust only two main parameters of adj1(10) (the K factor multiplying
the differential cross sections for hard parton-parton process) and adj1(34) (the
width of $p_T$ sampling). Other parameters are default. The adj1(10)	
and adj1(34) are fixed by fitting the ALICE measured $p_T$ spectra of $\pi^{+}+\pi^{-}$, $K^{+}+K^{-}$, $K^{0}_{S}$, 
$(D^{0}+\bar{D}^0)/2$ and $(D^{+}+D^{-})/2$ at mid-rapidity region in $pp$ collisions at
$\sqrt{s}$ = 7 TeV~\cite{ALICE:2015ial,ALICE:2017olh,ALICE:2020jsh}. The simulated	
$p_T$ spectra (with adj1(10)=0.8 and adj1(34)=0.46) and the
corresponding experimental data are shown in
Fig~\ref{fig_Tcs2870_pi_K_D_dN_dpT}. The simulated yields of $\pi^{+}$, $K^{-}$, $K^{0}_{S}$, 
$D^{0}$ and $D^{+}$ and the
corresponding experimental data are also presented in
Table~\ref{table:yields_exp_D_meson_Eng}.

\begin{table}[h]
	\caption{The yields of $\pi^{+}$, $K^{-}$, $K^{0}_{S}$, 
    $D^{0}$ and $D^{+}$ simulated by PACIAE model, compared with the ALICE data at mid-rapidity region in $pp$
    collisions at $\sqrt{s}$=7 TeV~\cite{ALICE:2015ial,ALICE:2017olh,ALICE:2020jsh}. Here the yields of $D^0$ and $D^+$ are prompt production (average of particles and antiparticles). }\label{table:yields_exp_D_meson_Eng}
	\centering
	\setlength{\tabcolsep}{8pt}
	\renewcommand{\arraystretch}{1.2}
	\begin{tabular}{ccc}
		\hline
		\hline
	  Particle & ALICE & PACIAE \\
		\hline
		$\pi^{+}$ & $2.26 \pm 0.10$ &  2.17  \\
	  $K^{-}$ & $0.286 \pm 0.016$ & 0.297  \\
        $K^{0}_{S}$ & $0.280 \pm 0.015 $ & 0.300  \\
		$D^0$ & $(6.83 \pm 0.77) \times 10^{-3}$ & $6.38 \times 10^{-3}$ \\
		$D^{+}$ & $(3.10 \pm 0.84) \times 10^{-3}$ & $2.99 \times 10^{-3}$ \\
		\hline
		\hline
	\end{tabular}
\end{table}

Then we use above generated final hadronic states to recompose four
sets of $T_{cs0}^{*}(2870)^{0}$ hadronic molecular state candidates by the
DCPC model using the constituent meson pairs of $D^{0}K^{0}_{S}$, $D^{+}K^{-}$, $D^{*+}K^{*-}$, and $D^{*0} \bar{K}^{*0}$, respectively. 

\begin{figure*}[!htbp]
\centering
\includegraphics[width=1.00\textwidth]{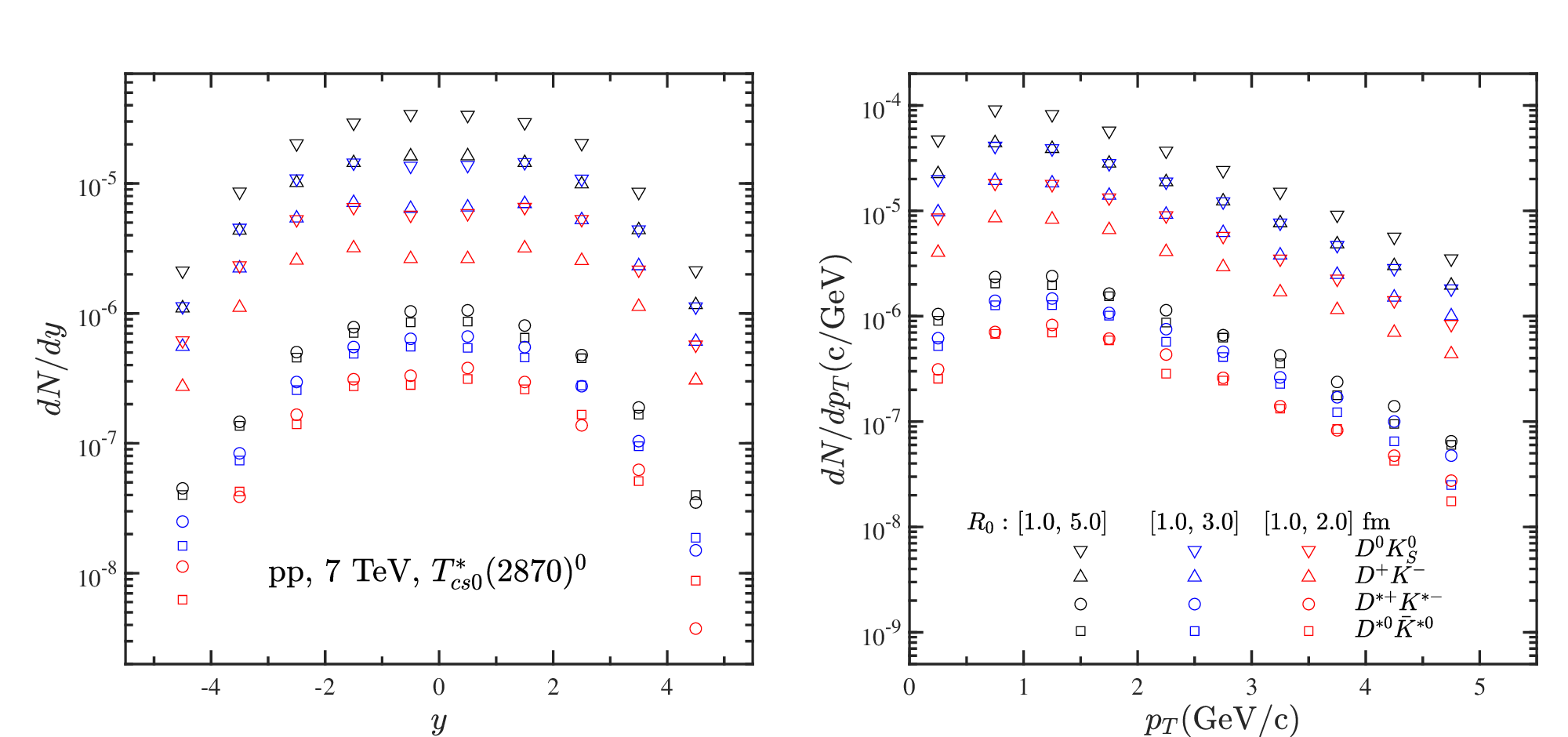}
\caption{ \centering{The simulated $y$ (left panel) and $p_T$ (right panel)
distributions of four $T_{cs0}^{*}(2870)^{0}$ candidates recombined from the $D^{0}K^{0}_{S}$ (triangles-down), the $D^{+}K^{-}$ (triangles-up), the $D^{*+}K^{*-}$
(circles), and the $D^{*0} \bar{K}^{*0}$ (squares) in $pp$ collisions at
$\sqrt{s}$ = 7 TeV are shown, respectively. The black, blue, and red markers refer
to results calculated in $R_0$ range of 1.0 fm $<R_0<$ 5.0 fm,
1.0 fm $<R_0<$ 3.0 fm, and 1.0 fm $<R_0<$ 2.0 fm, respectively.}
}\label{fig_Tcs2870_display_y_pT_state}
\end{figure*}

The simulated $y$ (left panel) and $p_T$ (right panel) distributions of four
$T_{cs0}^{*}(2870)^{0}$ candidates (recombined by $D^{0}K^{0}_{S}$, $D^{+}K^{-}$, $D^{*+}K^{*-}$, and $D^{*0} \bar{K}^{*0}$) in $pp$ collisions at $\sqrt{s}$ = 7 TeV are
given in Fig.~\ref{fig_Tcs2870_display_y_pT_state}. In this figure, the black,
blue, and red markers are the results calculated in the $R_0$ range of
1.0 $<R_0<$ 5.0 fm, 1.0 $<R_0<$ 3.0 fm, and 1.0 $<R_0<$ 2.0 fm, respectively.
The triangles-down, the triangles-up, the circles, and the squares refer to, respectively, the
results of the $D^0K^0_S$, $D^+K^-$, $D^{*+}K^{*-}$, and $D^{*0} \bar{K}^{*0}$ component meson
pairs. It is observed that the yield of $T_{cs0}^{*}(2870)^{0}$
composed of $D^{0}K^{0}_{S}$ is larger than the one composed of $D^{+}K^{-}$, 
which is consistent with the yields of the constituent particles of $T_{cs0}^*(2870)^0$ in Table~\ref{table:yields_exp_D_meson_Eng}. 
The yields of $T_{cs0}^{*}(2870)^{0}$
composed of $D^{*+}K^{*-}$ and $D^{*0} \bar{K}^{*0}$ are nearly equal, which is attributed to the similar yields of their constituent mesons $D^{*+}$ and $D^{*0}$ as reported in Ref.~\cite{LEBC-EHS:1988oic}.
The yields of $T_{cs0}^{*}(2870)^{0}$
composed of $D^{*+}K^{*-}$ and $D^{*0} \bar{K}^{*0}$ are lower than the former two. It can be understood by the fact that the productions of inclusive $D^+$ ($D^0$) and $K^-$ ($K^0_S$) are larger than those of $D^{*+}$ ($D^{*0}$) and $K^{*-}$ ($\bar{K}^{*0}$).
Meanwhile, the yield of $T_{cs0}^{*}(2870)^{0}$ increases with the increasing of the $R_0$ range from
1.0 fm $<R_0<$ 2.0 to 1.0 fm $<R_0<$ 3.0 fm, and to 1.0 fm $<R_0<$ 5.0 fm,
as expected.

\begin{figure*}[!htbp]
\centering
\includegraphics[width=1.00\textwidth]{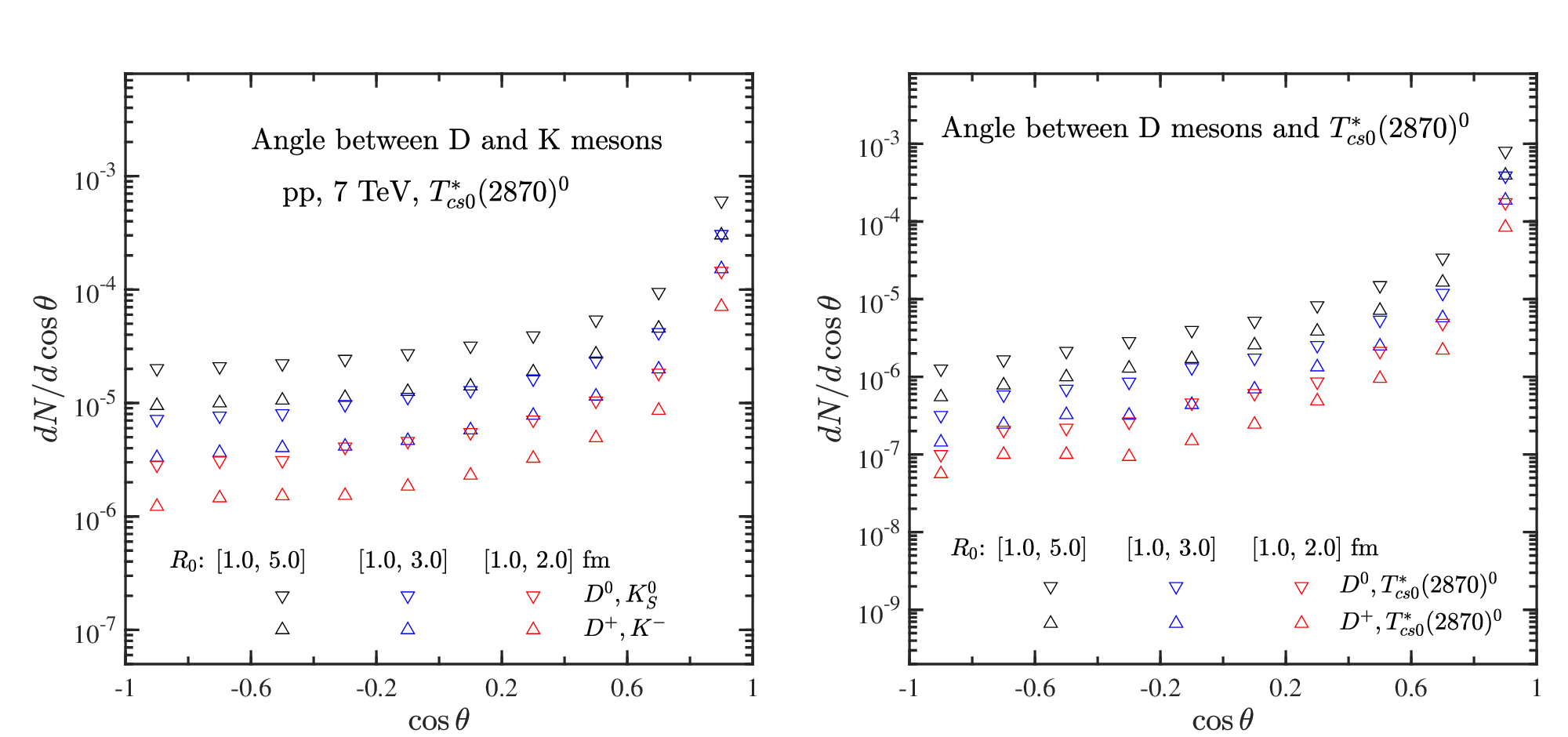}
\caption{ \centering{The angular distribution between two component mesons (left panel, where
triangles-down for ($D^0$, $K^0_s$) and triangles-up for ($D^+$, $K^-$)), as well as the one between $D$ component meson and $T_{cs0}^{*}(2870)^{0}$ (right panel, where
triangles-down for
($D^0$, $T_{cs0}^{*}(2870)^{0}$) and triangles-up for ($D^+$, $T_{cs0}^{*}(2870)^{0}$)) in $pp$ collisions at $\sqrt{s}$=7 TeV. The
black, blue, and red markers refer to the results calculated
in the $R_0$ range of 1.0 fm $<R_0<$ 5.0 fm, 1.0 fm $<R_0<$ 3.0 fm, and
1.0 fm $<R_0<$ 2.0 fm, respectively.} }\label{fig_Tcs2870_display_cos_state}
\end{figure*}

We also calculate the angular distribution between two component mesons
\begin{equation}
\cos \theta=\frac{ \vec{p}_1. \vec{p}_2 }{ | \vec{p}_1| | \vec{p}_2| }, \label{eq_cos_p12}
\end{equation}
where $\vec{p}_1$ ($\vec{p}_2$) denotes the momentum of $D$ ($K$)
component meson. Similarly, the angular distribution between $D$ component
meson and $T_{cs0}^{*}(2870)^{0}$ can be calculated. The above two angular
distributions are given in Fig.~\ref{fig_Tcs2870_display_cos_state}.
Note that the results of the angular distribution between $D^{*+}$ and $K^{*-}$, $D^{*0}$ and $\bar{K}^{*0}$, as well as $D^{*+}$ ($D^{*0}$) and $T_{cs0}^{*}(2870)^{0}$ are not shown here due to their immense computational demands.
In this figure, the black, blue, and red markers refer to the results calculated in
the $R_0$ range of 1.0 $<R_0<$ 5.0 fm, 1.0 $<R_0<$ 3.0 fm, and
1.0 $<R_0<$ 2.0 fm, respectively.  
The triangles-down and the triangles-up refer to, respectively, the results of the angle between $D^0$ and $K^0_S$,
as well as between $D^+$ and $K^-$
in the left panel. In the right panel, they refer to the results of the angle
between $D^0$ and $T_{cs0}^{*}(2870)^{0}$, as well as
$D^+$ and $T_{cs0}^{*}(2870)^{0}$, respectively.

In the left panel of Fig.~\ref{fig_Tcs2870_display_cos_state}
we observe the yield of $T_{cs0}^{*}(2870)^{0}$ is the largest (smallest) if
the angle between momentum vectors of $D$ and $K$ component meson pair is
close to $0^\circ$ ($\cos \theta$ = 1) (if the angle between momentum vectors
of $D$ and $K$ component meson pair is close to
$180^\circ$ ($\cos \theta$ = -1)).
It indicates the $T_{cs0}^{*}(2870)^{0}$ is easier to produce if the momentum
vectors of $D$ and $K$ meson pair are convergent, as expected. The similar
behavior is also observed in the right panel. The
Fig.~\ref{fig_Tcs2870_display_cos_state} shows also that the yield of
$T_{cs0}^{*}(2870)^{0}$ increases with increasing $R_0$ range from
1.0 fm $<R_0<$ 2.0 fm, to 1.0 fm $<R_0<$ 3.0 fm, and to
1.0 fm $<R_0<$ 5.0 fm.

\section{Summary}
\label{Summary}
In this letter, we successfully recombined four sets of
$T_{cs0}^{*}(2870)^{0}$ hadronic molecular state candidates based on the PACIAE model
simulated final hadronic state in $pp$ collisions at $\sqrt{s}$ = 7 TeV. We find that the yields
of four $T_{cs0}^{*}(2870)^{0}$ candidates follow the magnitude order of $D^{0}K^{0}_{S}$ $>$ $D^{+}K^{-}$ $>$ $D^{*+}K^{*-}$ $\sim$ $D^{*0} \bar{K}^{*0}$. We predict the $y$
distributions, the $p_T$ spectra, and the angular distributions for the four
sets of $T_{cs0}^{*}(2870)^{0}$ hadronic molecular state candidates, which are worthy
to be experimentally measured in future.

The extended studies for the $pp$ collisions at $\sqrt{s}$ = 8 and 13 TeV and
for the $p$-Pb and/or the Pb-Pb collisions at LHC energies are required.
Other observables, such as elliptic flow, etc. are also worthy to be
investigated.


%
%
%
%
%
%
%
%
%
%
%
%
\section*{Acknowledgements}
This work is supported by the National Natural Science
Foundation of China under Grants  No. 12375135. Y.L.Y. acknowledges the Continuous Basic
Scientific Research Project (Grant No. WDJC-2019-13). The work of W.C.Z. is supported
by the Natural Science Basic Research Plan in Shaanxi Province of
China(Program No. 2023-JCYB-012). H.Z. acknowledges the financial support from
Key Laboratory of Quark and Lepton Physics in Central China Normal University
under grant No. QLPL2024P01.	

%
%
%


\begin{thebibliography}{}
\bibitem{Zweig:1964ruk}
G.~Zweig,
An SU(3) model for strong interaction symmetry and its breaking. Version 1,
doi:10.17181/CERN-TH-401
\end{thebibliography}






\end{document}